\documentclass[conference]{IEEEtran}
\IEEEoverridecommandlockouts
\usepackage{cite}
\usepackage{amsmath,amssymb,amsfonts}
\usepackage{algorithmic}
\usepackage{graphicx}
\usepackage{textcomp}
\usepackage{xcolor}
\usepackage{balance}
\def\BibTeX{{\rm B\kern-.05em{\sc i\kern-.025em b}\kern-.08em
    T\kern-.1667em\lower.7ex\hbox{E}\kern-.125emX}}

\newcommand{\C}{{\cal C}}
\newcommand{\D}{{\cal D}}
\newcommand{\N}{{\cal N}}
\allowdisplaybreaks
\begin{document}

\title{Optimized Constellation Design for Two User Binary Sensor Networks Using NOMA\\

\thanks{The authors are with the Department of Mathematics and Statistics, Queen's University,
Kingston, Ontario, Canada (\{16ls53,takahara,fa\}@queensu.ca)).
This work was supported in part by NSERC of Canada.}
}

\author{Luca Sardellitti, Glen Takahara, Fady Alajaji}


\maketitle

\begin{abstract}
Data Fusion of wireless sensors is a common technique employed in many communication systems. This work focuses on incorporating the principles of non-orthogonal-multiple-access (NOMA) to optimize error performance directly in the choice of constellation design. 
More specifically, the problem of two sensor data fusion of a binary uniform source sent over a Gaussian multiple access channel via symmetric binary constellations is investigated. A so-called planar upper bound on the error probability is analytically derived. A constellation design is then obtained by establishing in closed form its rotation parameter that minimizes the upper bound. Simulation results show that the resulting constellations achieve a near identical performance as experimentally determined optimal constellations.

\end{abstract}


\section{Introduction}
There has been increased interest recently in the spectrally efficient non-orthogonal-multiple-access (NOMA) technique where multiple users send different data (via superposition) across a common channel (e.g., see~\cite{saito2013, ding17, vaezi2019interplay,makki2020survey,lin19,liu2022} and the references therein). In this work, the concept of sensor networks is combined with NOMA for the design of optimal symmetric binary constellations. Previous works such as \cite{mukho04} used functional processing of multiple sensor nodes' data before sending it over the channel to perform predictive error correction. 

Another common error mitigation approach is called data fusion. This is when multiple sensors simultaneously observe the same data source and then send their data over a shared channel, where a receiver is designed to decode the original source data. Previous research in this area has focused on optimizing the data fusion algorithm at 
the receiving node, while working with a fixed modulation scheme for sending the sensors' data. For example, \cite{ferrari14} fixes the constellation type to be binary phase-shift keying (BPSK), and \cite{jiang05} uses local likelihood ratio based on-off keying. 
This work instead focuses on finding the optimal constellation design 
so that the signals from the sensors can better reinforce each other and achieve a minimal error probability.
This is similar in concept to \cite{lin19} and \cite{weng18}, where the optimal constellation rotation angle is investigated. This work however considers a different problem since \cite{lin19} and \cite{weng18} deal with decoding two different sources sent over a multiple access channel, while the focus herein is on data fusion where a single source is decoded from two signals that are correlated to the source and sent via symmetric binary NOMA constellations over the multiple access channel. 

When analyzing constellation design, various possible configurations of parameters such as sensor correlation, noise and power constraints can affect the results of problem. 
The main contribution of this work is an upper bound on the optimal bit error rate (or error probability) which, when minimized to design the NOMA system's rotation parameter, is shown to be experimentally close to the simulated optimal error performance.
The problem setup, including the mathematical model, is described in Section~\ref{modelSec}. In Section~\ref{errorSection}, the system's error behaviour is analyzed in detail and an upper bound on the error probability based on planar decoding regions is established. Section~\ref{resultsSection} compares numerical results from the upper bound to the simulated optimal error performance. In Section~\ref{hignSNRSect}, it is shown that at high SNRs the error upper bound approaches the optimal error performance. Finally, future work directions are stated in Section~\ref{future-work}.

\section{System Model} \label{modelSec}
\subsection{Source and Sensors}
Let $X$ be a memoryless binary data source that is to be observed by a sensor network. For simplicity, the source is assumed to be uniformly distributed; i.e., $\text{Pr}(X = i)=1/2$, $i=0,1$. There are two sensors, $X_1$ and $X_2$ observing the source $X$, which are modelled as passing $X$ through two memoryless binary symmetric channels:
\begin{align}
    X_s &= X \oplus Z_s, \qquad \qquad s=1,2, 
\end{align}
where $Z_1$ and $Z_2$ are independent Bernoulli noise processes 
with means (or channel crossover probabilities) $\epsilon_1$ and $\epsilon_2$, respectively. To avoid uninteresting cases, sensor 1 is assumed to have stronger correlation to the original source $X$ than sensor~2: $0<\epsilon_1 < \epsilon_2<0.5$. Also let $P_1$ and $P_2$ denote the power constraints of the sensors (each sensor has its own power allotment, as opposed to having a common power constraint on the entire network).
The sensors, unable to communicate with each other, encode their data independently using symmetric binary constellations. The constellations for the sensors are parameterized as follows: $\C_1 = \{c_{0,1}, c_{1,1}\} = \{-\sqrt{P_1},\; \sqrt{P_1}\}$ and $\C_2 = \{c_{0,2}, c_{1,2}\} = \{-\sqrt{P_2}e^{j\theta},\; \sqrt{P_2}e^{j\theta}\}$, where $j$ is the imaginary unit and $\theta \in [0,\frac{\pi}{2}]$ is the rotation between the two constellations. For $i\in \{0,1\}$, $s\in \{1,2\}$  $c_{i,s} \in \C_s$ denotes the constellation point assigned to $X_s = i$.

\subsection{Channel Model}
The two sensor signals are superimposed and sent through a Gaussian multiple access channel (GMAC) with noise power $N_0$. The received signal $R$ is given by
\begin{equation}
    R = S_1 + S_2 + Z
\end{equation}
where $S_1\in \C_1$, $S_2 \in \C_2$ are each sensor's chosen constellation point, and $Z$ is 
a complex (bivariate) Gaussian noise variable
with independent zero mean components of equal variance given by $\frac{N_0}{2}$. It is also assumed that $Z$ is independent of the sensor signals $S_1$ and $S_2$. Due to the superposition of the sensor signals, the overall signal $S_1+S_2$ sent over the channel can be represented as a point in the combined constellation of $\C_1$ and $\C_2$, given by
\begin{equation}
    \C = \{c_1 + c_2 \:|\: c_1 \in \C_1,\: c_2 \in \C_2\}.
\end{equation}
This work's objective is then to design optimal symmetric binary NOMA constellations in the sense of achieving the smallest possible error probability. Since sensor~1's constellation $\C_1$ is fixed as BPSK, it is sufficient to optimize constellation $\C_2$ for sensor~2, which is equivalent to finding the optimal rotation angle $\theta$. This problem is tackled by establishing and optimizing a tight upper bound on the error probability.

\section{Error Analysis and Constellation Design} \label{errorSection}

\subsection{Decoding Regions}
\begin{figure}[htbp]
\centerline{\includegraphics[scale=0.4]{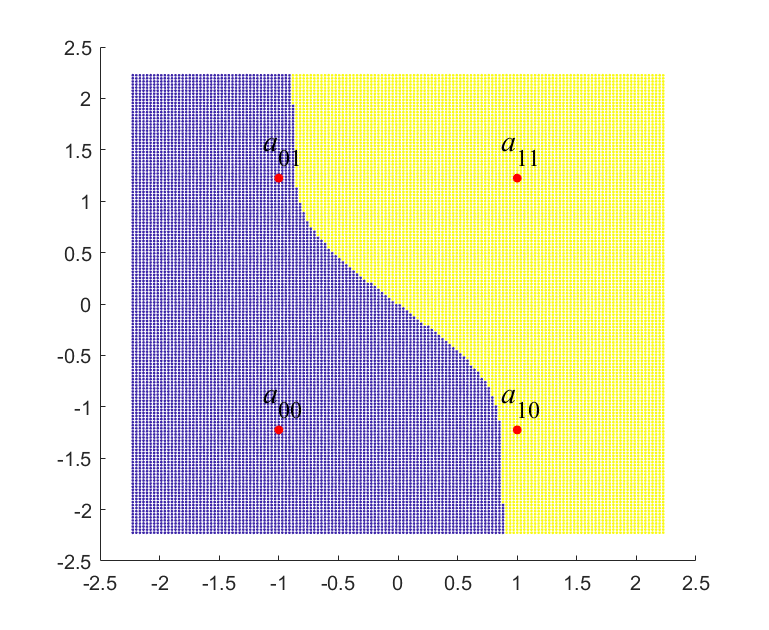}}
\vspace{-0.1in}
\caption{Decision regions for $\theta = \frac{\pi}{2}$ (yellow region is $\D_1$). The red points represent the superimposed constellation points $a_{lm} \in \mathcal{C}$.}
\label{fig1}
\end{figure}

To recover the original data source, maximum-likelihood (ML) decoding is used (which is optimal as the source is uniform). For a GMAC received signal $r\in \mathbb{C}$, the decoded symbol is determined as follows:
\begin{align}
    \hat{x}(r) &= \underset{i\in\{0,1\}}{\arg\max}\; f(R = r\;|\;X = i) \nonumber \\
   &= \underset{i\in\{0,1\}}{\arg\max}\; \sum_{(l,m)\in\{0,1\}^2}p_{lm|i}e^{-\frac{|r-a_{lm}|^2}{N_0}}
\end{align}
where $f$ is the conditional probability density function (pdf), $p_{lm|i} \triangleq\text{Pr}(X_1 = l, X_2 = m | X = i)$, and $a_{lm} \in \C$ denotes the superimposed constellation symbol associated with $X_1 = l$ and $X_2 = m$. Note that the conditional probabilities $p_{lm|i}$ can be expressed in terms of the sensor crossover probabilities:
\begin{align}
\nonumber
    p_{11|0} &= p_{00|1} = \epsilon_1\epsilon_2, \quad
    p_{00|0} = p_{11|1} = (1-\epsilon_1)(1-\epsilon_2) \\ \nonumber    
    p_{01|0} &= p_{10|1} = (1-\epsilon_1)\epsilon_2,\quad
    p_{10|0} = p_{01|1} = \epsilon_1(1-\epsilon_2).
\end{align}
To decode $r$, the complex plane is partitioned into two decoding (or decision) regions, $\D_0$ and $\D_1=\D_0^c$, where $\D_i = \{r\;|\;\hat{x}(r) = i\}$, $i=0,1$.
By taking advantage of symmetries, $\D_1$ can be expressed explicitly as function of the rotation angle $\theta$ between constellations $\C_1$ and $\C_2$ as follows:
\begin{align*}
    \D_1(\theta) &= \Bigg\{r\;\Big|\; \sum_{(l,m)\in\{0,1\}^2}p_{lm|1}e^{-\frac{|r-a_{lm}|^2}{N_0}} >    \\
     & \qquad \qquad \qquad \qquad \sum_{(l,m)\in\{0,1\}^2}p_{lm|0}e^{-\frac{|r-a_{lm}|^2}{N_0}} \Bigg\} \\
    & = \hspace{-0.02in} \Bigg\{r\;\Big|\; (p_{11|1} - p_{11|0})e^\frac{-|a_{11}|^2}{N_0}\Big(e^\frac{2\text{Re}(ra_{11}^*)}{N_0} - e^\frac{-2\text{Re}(ra_{11}^*)}{N_0}\Big)  \\
    & \quad >(p_{10|1} - p_{01|0})e^\frac{-|a_{01}|^2}{N_0}\Big(e^\frac{2\text{Re}(ra_{01}^*)}{N_0}-e^\frac{-2\text{Re}(ra_{01}^*)}{N_0}\Big)  \Bigg\}  \\ 
    & = \Bigg\{r\;\Big|\; K_1(\theta)\sinh\Big(\frac{2\text{Re}(rc_{1,1}^*) + 2\text{Re}(rc_{1,2}^*)}{N_0}\Big)  \\ 
    & \qquad\quad > K_0(\theta)\sinh\Big(\frac{-2\text{Re}(rc_{1,1}^*) + 2\text{Re}(rc_{1,2}^*)}{N_0}\Big)  \Bigg\}\\
   &= \Bigg\{r\;\Big|\; \tanh{A(r)} > \tanh{B(r,\theta)}\frac{K_0(\theta)-K_1(\theta)}{K_0(\theta)+K_1(\theta)}\Bigg\} 
\end{align*}
where $^*$ denotes complex conjugation,
\begin{align} \nonumber
    A(r) = \frac{2\text{Re}(rc_{1,1}^*)}{N_0}, \quad&
    B(r, \theta) = \frac{2\text{Re}(rc_{1,2}^*)}{N_0}, \\ \nonumber
    K_0(\theta) = (\epsilon_2 - \epsilon_1)e^\frac{-|a_{01}|^2}{N_0}, \quad&
    K_1(\theta) = (1-\epsilon_1 - \epsilon_2)e^\frac{-|a_{11}|^2}{N_0},
\end{align}
and the identity $|a+b|^2 = |a|^2 + 2\text{Re}(ab^*)+ |b|^2$ was used along with the facts that $a_{11} = -a_{00}$ and $a_{01} = -a_{10}$. 
Note also that the constellation points $c_{1,2}$, $a_{01}$ and $a_{11}$ are all a function of~$\theta$. An example of regions $\D_0$ and $\D_1$ is depicted in Fig.\ref{fig1} for $N_0 = 1$, $P_1 = 1$, $P_2 =1.5$, $\epsilon_1 = 0.15$, $\epsilon_2 = 0.17$ and $\theta = \frac{\pi}{2}$. 

\subsection{Error Upper Bound for Planar Decision Regions}
The system's error probability (under optimal ML decoding and uniform source $X$) can be expressed as 
\begin{equation}
P_e = \text{Pr}(R\in \D_0 \;|\; X = 1). \label{orig-pe}
\end{equation}
Note that $P_e=P_e(\theta)$, i.e., it is a function of $\theta$, and the boundary between regions $\D_0$ and $\D_1$ is given by
\begin{align} \label{boundaryCondition}
    \tanh{A(r)}= \tanh B(r,\theta)\frac{K_0(\theta)-K_1(\theta)}{K_0(\theta)+K_1(\theta)}.
\end{align}
An upper bound on $P_e(\theta)$ in~\eqref{orig-pe} is next obtained by restricting the decision regions to be {\em planar}, i.e., they are the left and right half planes: $$\D_1^c=\D_0 = \D_{0,\text{planar}} \triangleq \{r \;|\; \text{Re}(r) \le 0\}.$$ Thus
\begin{align} \nonumber \label{errorProbSuperposition}
    P_e(\theta) & = \text{Pr}(R\in \D_0 \;|\; X = 1) \\ \nonumber
    & \le \text{Pr}(R\in \D_{0,\text{planar}} \;|\; X = 1) \\ \nonumber
    & = \text{Pr}(\text{Re}(R) \leq 0 \;|\; X = 1) \\ \nonumber
    & = \sum_{(l,m)\in\{0,1\}^2}p_{lm|1}\text{Pr}(\text{Re}(R) \leq 0 | X_1 = l, X_2 = m) \\ 
    & = \epsilon_1 + (1 - \epsilon_1 - \epsilon_2)Q\left(\frac{\sqrt{P_1} + \sqrt{P_2}\cos(\theta)}{\sigma}\right) \nonumber \\ & \qquad \qquad + (\epsilon_2 - \epsilon_1)Q\left(\frac{\sqrt{P_1} - \sqrt{P_2}\cos(\theta)}{\sigma}\right) \nonumber \\ 
    &\triangleq P_e^{\text{ub}}(\theta),
\end{align}
where $\sigma=\sqrt{N_0/2}$ denotes the noise standard deviation, $Q(x)=\frac{1}{\sqrt{2\pi}} \int_x^\infty e^{-t^2/2} \, dt$ is the Gaussian tail function and the penultimate equality holds since for a fixed $(X_1,X_2) =(l,m)$, 
the real part of the received signal, 
$\text{Re}(R)$, is normal (Gaussian) with mean $\text{Re}(a_{lm})$ and variance~$N_0/2$.


\subsection{Existence of Planar Decision Regions}\label{planar-exist}
With the error upper bound $P_e^{\text{ub}}(\theta)$ obtained in~\eqref{errorProbSuperposition} for planar decision regions, the objective is to minimize it over the constellation angle $\theta$ and hence construct an optimized constellation $\C_2$. 
But it is first shown that such planar decision regions indeed exist; that is,
there is always a value of $\theta$ that will result in the regions $\D_0$ and $\D_1$ being the left and right half planes, respectively. In order for the decision boundary to be exactly on the imaginary axis it is required that for all $r$ such that $\text{Re}(r) = 0$, \eqref{boundaryCondition} must hold. Since $\tanh A(r) = 0$ for $\text{Re}(r) = 0$, and $\tanh B(r, \theta) \neq 0$ for $\text{Im}(r) \neq 0$ and $\theta \neq 0$, \eqref{boundaryCondition} yields that
\begin{align} \nonumber
    &\hspace{-0.25in} \frac{K_0(\theta)-K_1(\theta)}{K_0(\theta)+K_1(\theta)}= 0 \nonumber \\ &\implies 
    K_0(\theta) = K_1(\theta) \nonumber \\ \nonumber
    &\implies (1-\epsilon_1 - \epsilon_2)e^{\frac{-|a_{11}|^2}{N_0}}  = (\epsilon_2 - \epsilon_1)e^{\frac{-|a_{01}|^2}{N_0}} \\
    &\implies \theta = \cos^{-1}(pcf) \label{opt-theta}
\end{align}
for $0\le pcf\le 1$, where 
\begin{equation}
    pcf \triangleq \frac{N_0}{4\sqrt{P_1P_2}}\ln\left(\frac{1-\epsilon_2 - \epsilon_1}{\epsilon_2 - \epsilon_1}\right) \label{eq-pcf}
\end{equation}
is called the {\em power-correlation-factor}.
We remark that the $pcf$ given in~\eqref{eq-pcf} may be larger than 1, in which case \eqref{opt-theta} does not hold. When $pcf \leq 1$, it is easy to verify that $\theta$ in~\eqref{opt-theta} yields that the decision regions are exactly the left and right halves of the plane (note that $pcf>0$ since $0<\epsilon_1<\epsilon_2<1/2$).

When $pcf > 1$ it is next shown that for $\theta = 0$, the resulting decision regions are also the left and right halves planes. First note that when $\theta = 0$, the constants $K_0$ and $K_1$ reduce~to
\begin{align}
    K_0(0) &= (\epsilon_2 - \epsilon_1)e^\frac{-P_1 - P_2 + 2\sqrt{P_1P_2}}{N_0} \label{eq-k0} \\
    K_1(0) &= (1 - \epsilon_2 - \epsilon_1)e^\frac{-P_1 - P_2 - 2\sqrt{P_1P_2}}{N_0}. \label{eq-k1}
\end{align}
Now in light of \eqref{eq-k0} and \eqref{eq-k1}, the condition $pcf > 1$ implies 
\begin{align} \nonumber
    \frac{N_0}{4\sqrt{P_1P_2}}&\ln\Big(\frac{1-\epsilon_2 - \epsilon_1}{\epsilon_2 - \epsilon_1}\Big) > 1 \\ \nonumber
    &\implies \frac{1-\epsilon_2 - \epsilon_1}{\epsilon_2 - \epsilon_1} > e^\frac{4\sqrt{P_1P_2}}{N_0} \\ \nonumber
    & \implies (1-\epsilon_2 - \epsilon_1)e^\frac{-2\sqrt{P_1P_2}}{N_0} > (\epsilon_2 - \epsilon_1)e^\frac{2\sqrt{P_1P_2}}{N_0} \\ \nonumber
    &\implies K_0(0) < K_1(0) \\ 
    &\implies \frac{K_0(0)-K_1(0)}{K_0(0)+K_1(0)} < 0. \nonumber
\end{align}
Finally, noting that $\text{sgn}(\tanh{A(r)}) = \text{sgn}(\tanh{B(r,0)})$ for any $r$, it directly follows that the resulting decoding regions are the left and right halves of the complex plane.

\subsection{Optimizing the Planar Error Bound}
To optimize \eqref{errorProbSuperposition} over $\theta$, first observe that the only dependence on $\theta$ is in a cosine function, so the function will be periodic and even. Hence it is sufficient to minimize it over the interval $\theta \in [0,\pi]$. Since the expression is a bounded, differentiable function of $\theta$, finding the smallest critical point or endpoint will solve the minimization problem. Solving for the critical points is done as follows, recalling that the derivative of the Gaussian tail function is $\frac{d}{dx}Q(x) = -f_\N(x)$, where $f_\N$ is the pdf of a standard normal random variable:
\begin{align} \label{errorDerivative}
    \frac{d}{d\theta}&P_e^{\text{ub}}(\theta) \nonumber \\ & \hspace{-0.05in} = (1 - \epsilon_1 - \epsilon_2)f_\N\Big(\frac{\sqrt{P_1} + \sqrt{P_2}\cos(\theta)}{\sigma}\Big)\Big(\frac{\sin\theta\sqrt{P_2}}{\sigma}\Big) \nonumber \\
    & \hspace{-0.05in} - (\epsilon_2 - \epsilon_1)f_\N\Big(\frac{\sqrt{P_1} - \sqrt{P_2}\cos(\theta)}{\sigma}\Big) \Big(\frac{\sin\theta\sqrt{P_2}}{\sigma}\Big).  
\end{align}
From \eqref{errorDerivative}, it is clear that if $\sin(\theta) = 0$, it results in a critical point. This is equivalent to $\theta \in \{0,\pi\}$, which takes care of the endpoints as well. Now, assuming $\sin(\theta) \neq 0$, and solving for when the above derivative is 0 yields: 
\begin{align} 
    &\frac{(1-\epsilon_1 - \epsilon_2)}{(\epsilon_2 - \epsilon_1)}e^{-\frac{(\sqrt{P_1} + \sqrt{P_2}\cos(\theta))^2}{2\sigma^2}}  = e^{-\frac{(\sqrt{P_1} - \sqrt{P_2}\cos(\theta))^2}{2\sigma^2}}  \nonumber \\
    &\implies (1-\epsilon_1 - \epsilon_2)e^{\frac{-2\cos(\theta)\sqrt{P_1P_2}}{N_0}} = (\epsilon_2 - \epsilon_1)e^{\frac{2\cos(\theta)\sqrt{P_1P_2}}{N_0}}  \nonumber \\
    &\implies \theta = \cos^{-1}(pcf) \label{opt-theta2}
\end{align}
where the above critical point exists if and only if $pcf \leq 1$. Hence the error bound expression has 2 or 3 critical points depending on the value of $pcf$. It  is however easy to verify that $P_e^{\text{ub}}(0) < P_e^{\text{ub}}(\pi)$ using the fact that $\cos(0) = -\cos(\pi) = 1$:
\begin{align} \nonumber
    P_e^{\text{ub}}(0)-P_e^{\text{ub}}(\pi) &= (1-\epsilon_1-\epsilon_2)(Q_1-Q_2) 
     \\ \nonumber & \qquad \quad 
    + (\epsilon_2 - \epsilon_1)(Q_2-Q_1) \\ \nonumber
    & = (1-2\epsilon_2)(Q_1-Q_2)  \nonumber
    < 0
\end{align}
where $Q_1 
\triangleq Q(\frac{\sqrt{P_1} + \sqrt{P_2}}{\sigma})$, $Q_2 \triangleq Q(\frac{\sqrt{P_1} - \sqrt{P_2}}{\sigma})$ and the inequality follows since $Q_1 < Q_2$ (as the $Q$-function is decreasing).
All that remains is to compare $P_e(0)$ and $P_e(\cos^{-1}(pcf))$ (if it exists). It suffices to show in this case that the derivative of the error expression in \eqref{errorDerivative} is less than zero for each $\theta \in (0,\cos^{-1}(pcf))$. First note that $\theta \in (0,\cos^{-1}(pcf)) \implies \cos(\theta) \in (pcf,1)$, and hence, recalling the expression of $pcf$ in~\eqref{eq-pcf}, the following holds for these $\theta$ values:
\begin{align} \label{errorInequality}
    (1-\epsilon_1-&\epsilon_2)e^{\frac{-2\cos(\theta)\sqrt{P_1P_2}}{N_0}} - (\epsilon_2-\epsilon_1)e^{\frac{2\cos(\theta)\sqrt{P_1P_2}}{N_0}}  \nonumber \\
    &< (1-\epsilon_1-\epsilon_2)e^{\frac{-2pcf\sqrt{P_1P_2}}{N_0}} - (\epsilon_2-\epsilon_1)e^{\frac{2pcf\sqrt{P_1P_2}}{N_0}} \nonumber \\
    &= 0.
\end{align}
Thus, for $\theta \in (0,\cos^{-1}(pcf))$, ${d}P_e^{\text{ub}}(\theta)/d\theta$ in \eqref{errorDerivative} satisfies 
\begin{align} \nonumber
    \frac{d}{d\theta}&P_e^{\text{ub}}(\theta) = \frac{\sin\theta\sqrt{P_2}}{\sigma}\Big((1 - \epsilon_1 - \epsilon_2)e^{-\frac{(\sqrt{P_1} + \sqrt{P_2}\cos(\theta))^2}{N_0}} \\ \nonumber 
    &\qquad \qquad \quad - (\epsilon_2 - \epsilon_1)e^{-\frac{(\sqrt{P_1} - \sqrt{P_2}\cos(\theta))^2}{N_0}}\Big) \\ \nonumber
    &= \frac{\sin\theta\sqrt{P_2}}{\sigma}e^{\frac{P_1+\cos^2(\theta)P_2}{N_0}}((1 - \epsilon_1 - \epsilon_2)e^{-\frac{2\cos(\theta)\sqrt{P_1P_2}}{N_0}} \\ \nonumber 
    &\qquad \qquad \quad - (\epsilon_2 - \epsilon_1)e^{\frac{2\cos(\theta)\sqrt{P_1P_2}}{N_0}}) \\ \nonumber
    & < 0
\end{align}
where the last inequality holds by~\eqref{errorInequality} and the fact that $\sin(\theta) > 0$ for all $\theta \in (0,\cos^{-1}(pcf))$. Hence the smallest critical point (and minimum) of the planar error bound equals $\theta = \cos^{-1}(pcf)$, or $\theta = 0$ if $pcf > 1$.

\subsection{Least Planar Upper Bound}
The implication of the above results is that the smallest planar upper bound over the constellation rotation parameter $\theta$ is achieved at
\begin{equation}
    \theta_{ub}^* = \cos^{-1}(\min\{pcf,1\}) \label{opt-theta3}
\end{equation}
and, upon substituting \eqref{opt-theta3} into~\eqref{errorProbSuperposition}, is given by
\begin{align} \label{errorProbOptimal}
    \hspace{-0.07in} P_e^{\text{ub}}(\theta_{ub}^*)& = \epsilon_1 + (1 - \epsilon_1 - \epsilon_2)Q\left(\frac{\sqrt{P_1} + \sqrt{P_2}\min\{pcf,1\}}{\sigma}\right) \nonumber \\ & \quad + (\epsilon_2 - \epsilon_1)Q\left(\frac{\sqrt{P_1} - \sqrt{P_2}\min\{pcf,1\}}{\sigma}\right). 
\end{align}
Note that the $\theta$ in~\eqref{opt-theta3} minimizing $P_e^{\text{ub}}(\theta)$ is identical to the one derived in Section~\ref{planar-exist} yielding planar decision regions. 

\section{Numerical and Simulation Results} \label{resultsSection}
It is next demonstrated that this upper bound performs very well experimentally at any signal-to-noise ratio (SNR). This is achieved by comparing the optimal upper bound in~\eqref{errorProbOptimal} with an experimentally determined optimal error probability. The system's SNR is calculated as a geometric average, i.e., $\text{SNR} = \frac{\sqrt{P_1P_2}}{N_0}$, and it is reported in dB (i.e., $\text{SNR (dB)} = 10\log_{10}(\text{SNR})$). The process for generating the experimental data is as follows. For each SNR and $\theta$ value, 
$n=30$ trials, each consisting of $N=100,000$ independent source bits being sent through the channel, were simulated. 
The SNR values, ranging from -10 to 20 dB, are listed in Table~\ref{errorSNRTheta}, and 100 $\theta$ values, each equally spaced in $[0,\frac{\pi}{2}]$, were simulated.

\begin{figure}[hbtp]
\centerline{\includegraphics[scale=0.35]{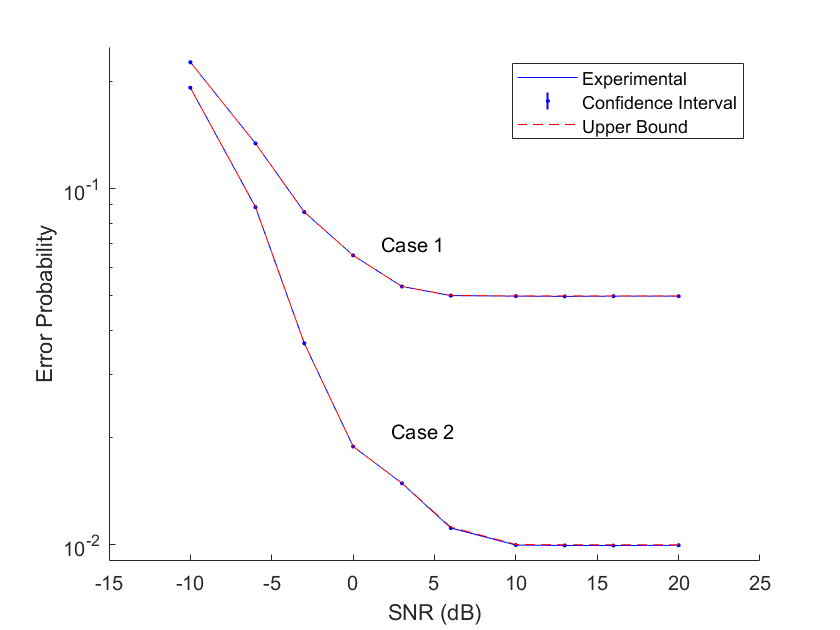}}
\vspace{-0.08in}
\caption{Optimal error probability for varying SNR values.}
\label{ErrorProbVsSNR}
\end{figure}

The ML decoding rule was applied to the simulated data and the error rate for each trial was calculated for each $\theta$. The error probability was estimated for each $\theta$ by averaging the 30 trial results. Then, after taking a moving average over $\theta$ of the estimated error probabilities, the minimum value with respect to $\theta$ was calculated, and the respective $\theta$ value which achieves this minimum was recorded. This is how $\theta^*_{exp}$ and $P_e(\theta^*_{exp})$ are presented in Table~\ref{errorSNRTheta} and Fig.~\ref{ErrorProbVsSNR}, respectively. Results for $\epsilon_1 = 0.05$, $\epsilon_2 = 0.1$, $P_1 = 2$ and $P_2 = 1$ (referred to as Case~1) and $\epsilon_1 = 0.01$, $\epsilon_2 = 0.02$, $P_1 = 1$ and $P_2 = 2$ (referred to as Case~2) are given in Table~\ref{errorSNRTheta}. 
\begin{table}[htbp]
\vspace{-0.07in}
\caption{Optimal $\theta$ values (in radians) minimizing  error probability}
\label{errorSNRTheta}
\begin{center}
\vspace{-0.15in}
\begin{tabular}{|c|c|c|c|c|c|c|}
\hline
&\multicolumn{2}{|c|}{Case 1} & \multicolumn{2}{|c|}{Case 2}\\
\hline
SNR (dB)&$\theta_{ub}^*$&$\theta_{exp}^*$&$\theta_{ub}^*$&$\theta_{exp}^*$ \\
\hline
-10 & 0 & 0.016  & 0 & 0 \\
-6 & 0 & 0 & 0 & 0.015 \\
-3 & 0 & 0.079 & 0 & 0 \\
0 & 0.784 & 0.777 & 0 & 0.143 \\
3 & 1.208 & 1.269 & 0.960 & 0.841 \\
6 & 1.392 & 1.491 & 1.279 & 1.079 \\
10 & 1.50 & 1.238 & 1.456 & 0.762 \\
13 & 1.535 & 0.809 & 1.513 & 0.793 \\
16 & 1.553 & 0.539 & 1.542 & 1.174 \\
20 & 1.563 & 0.190 & 1.559 & 0.365 \\
\hline
\end{tabular}
\end{center}
\end{table}
From Table~\ref{errorSNRTheta}, it can be seen that at low SNRs, the $\theta ^{\star}_{ub}$ values minimizing the error upper bound $P_e^{\text{ub}}(\theta)$ and the $\theta ^{\star}_{exp}$ values minimizing the simulated true error probability $P_e(\theta)$ are nearly identical.
This is not however the case at high SNRs; this discrepancy is attributed to the system's high SNR behaviour (analyzed in Section~\ref{hignSNRSect}), where it is shown that both $P_e^{\text{ub}}(\theta)$ and $P_e(\theta)$ approach the {\em same constant} ($\epsilon_1$) that is independent of $\theta$. Hence this explains why the optimal $\theta ^{\star}_{exp}$ values are hard to obtain accurately at high SNRs. However, this is not of practical interest since the optimal error probability is not sensitive to $\theta$ in the high SNR regime.
Furthermore, the optimal $\theta ^{\star}_{ub}$ values in Table~\ref{errorSNRTheta} are accurate for large SNRs and approach the theoretical limit of $\pi/2$ as shown in Section~\ref{hignSNRSect}. 

Finally, note that for any SNR, it is the
resulting optimal error values that are of primary interest, and these can be seen in Fig.~\ref{ErrorProbVsSNR}. The error bars show the $95\%$ confidence intervals for the optimal bit error rate and are calculated using the standard deviation of the minimum values over each of the $n$ trials.
From Fig.~\ref{ErrorProbVsSNR} note that, for the entire SNR range, the error probability graphs are nearly identical and the experimental error bars overlap with the upper bound at most data points. It is also observed that at high SNRs, although the optimal $\theta$ values do not line up very well, the resulting optimal error values at high SNR are still very close together.


\section{High SNR Analysis}\label{hignSNRSect}
For this analysis it is defined that high SNR means $N_0 \rightarrow 0$. This is a reasonable assumption since each of the individual sensor SNRs should be growing at similar rates, and one sensor should not have infinitely more power than the other.

\subsection{Upper Bound Analysis}
Taking the limit as $N_0\to 0$ of the power-correlation-factor's expression in~\eqref{eq-pcf} directly gives that 
\begin{align} \label{limitPcf} \nonumber
    \lim_{N_0\rightarrow0} pcf &= \lim_{N_0\rightarrow0} \frac{N_0}{4\sqrt{P_1P_2}}\ln\Big(\frac{1-\epsilon_2 - \epsilon_1}{\epsilon_2 - \epsilon_1}\Big)=0. \nonumber
\end{align}
This directly implies using~\eqref{opt-theta3} that
$\lim_{N_0\rightarrow0} \theta^*_{\text{ub}} = \frac{\pi}{2}$, which in turn yields using~\eqref{errorProbOptimal} that
\begin{align} \nonumber
\lim_{N_0\rightarrow0}P_e^{\text{ub}}(\theta_{ub}^*)& = 
\epsilon _1\nonumber 
\end{align}
since $\lim_{x\rightarrow\infty}Q(x)=0$. The above results show that the optimal upper bound constellation approaches {\em orthogonal signaling}, and that $P_e^{\text{ub}}(\theta_{ub}^*)$ approaches the error probability incurred by
just sending the signal with more correlation to the original source (i.e., just sending $X_1$).

\subsection{Asymptotic Optimality}
\begin{figure}[htbp]
\centerline{\includegraphics[scale=0.35]{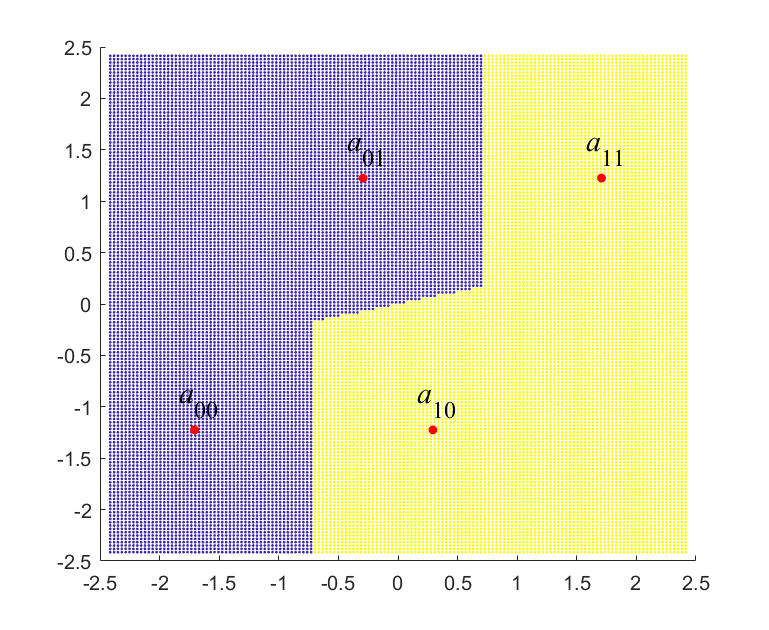}}
\vspace{-0.1in}
\caption{High SNR decision regions for $\theta = \frac{\pi}{3}$, $P_1 = 1$, $P_2 = 2$ (yellow region is $\D_1$).}
\label{HighSNRDecisionRegion}
\vspace{-0.1in}
\end{figure}

It is next pointed out that the upper bound $P_e^{\text{ub}}(\theta_{ub}^*)$ approaches the true minimum error probability $P_e(\theta^*)$ in the high SNR regime. Due to space limitations, this result is not proved rigorously, but it is instead explained following an intuitive argument. First note that for any fixed set of parameters (including $\theta$) the decision regions in the high SNR regime approach the nearest neighbour decoding regions for the sensor with more correlation to the original source. That is, the region $\D_1$ approaches the union of the two nearest neighbour decoding regions for the points associated with $X_1 = 1$; this is illustrated in~Fig.~\ref{HighSNRDecisionRegion}.
Note that the boundary of these limiting decision regions are combinations of straight lines. It can be verified that the two vertical lines occur at $x = \pm\sqrt{P_2}\cos(\theta)$ and the diagonal line is described by the equation $y = \frac{\sqrt{P_1} - \sqrt{P_2}\cos(\theta)}{\sqrt{P_2}\sin(\theta)}x$. The only exception to this is at $\theta = 0$, where there is no diagonal line, and instead there is a third vertical line at $x = 0$. Also observe that there is no dependence on $\epsilon_1$ or $\epsilon_2$ in the high SNR decision regions.

As SNR grows without bound, the probability that any sent constellation point exits its decision region approaches zero. Hence the decoding rule at high SNRs becomes $\hat{X} = X_1$. This immediately implies that the error probability at any $\theta$ approaches $P(X_1 \neq X) = \epsilon_1$. Hence 
$$\lim_{N_0\to 0}P_e(\theta^*)= \epsilon_1=\lim_{N_0\to 0}P_e^{\text{ub}}(\theta_{ub}^*)$$ 
and $P_e^{\text{ub}}(\theta_{ub}^*)$ is asymptotically optimal.

\section{Future Work}\label{future-work}
Given the experimental results showing strong correlation between the optimal upper bound and the overall optimal error performance, the first natural extension of this work is to prove that this upper bound is indeed the optimal value, or show conditions for when it is not the case, if there are any edge cases not herein considered.

Furthermore, the source distribution was taken to be uniformly distributed and the binary constellations were restricted to be symmetric. This problem could be generalized to non-uniform binary sources, or by adding more constellation parameters to allow non-symmetrical constellations, allowing potentially for better error performance.


\balance
\bibliographystyle{IEEEtran}
\bibliography{references}

\end{document}